\documentclass[11pt,twoside]{article}
\usepackage{asp2004}
\usepackage{psfig}
\usepackage{epsf}
\usepackage{graphics}
\usepackage{lscape}
\def\edcomment#1{\iffalse\marginpar{\raggedright\sl#1\/}\else\relax\fi}
\reversemarginpar

\newcommand{\astroe}{{\sl Astro-E2\/}}
\newcommand{\xmm}{{\sl XMM-Newton\/}}
\newcommand{\chandra}{{\sl Chandra\/}}
\newcommand{\sax}{{\sl BeppoSAX\/}}
\newcommand{\xte}{{\sl RXTE\/}}
\newcommand{\asca}{{\sl ASCA\/}}
\newcommand{\euve}{{\sl EUVE\/}}
\newcommand{\rosat}{{\sl ROSAT\/}}
\newcommand{\ginga}{{\sl Ginga\/}}

\newcommand{\einstein}{{\sl Einstein\/}}

\begin{document}
\title{Recent Advances in X-ray Observations of Cataclysmic Variables}
\author{K. Mukai}
\affil{NASA/Goddard Space Flight Center and Universities Space Research
Association, \\ Code 662, Greenbelt MD 20771, USA}

\begin{abstract}

A personal selection of noteworthy X-ray results on CVs are presented,
with emphasis on \xmm\ and \chandra\ observations.  Progressing roughly
from broad-band view to narrow-band, high spectral resolution studies,
I summarize: the energy balance of polars; X-ray confirmation of IPs;
eclipses in non-magnetic CVs; search for magnetism in ``non-magnetic''
CVs; multi-temperature plasma emission from the boundary layer;
complex absorption in magnetic CVs; temperature and density diagnostics;
and X-ray radial velocity studies.

\end{abstract}
\thispagestyle{plain}

\section{The X-ray Band}

Over the past 25 years or so, X-ray observations have played a major role 
in the research into CVs.  It would only be a slight overstatement to say
that the class of magnetic CVs owes its existence to X-ray astronomy,
as a high fraction of magnetic CVs are discovered and/or confirmed through
X-ray observations.  The role X-rays have played for non-magnetic CVs
is less central, but nevertheless important.  This is because, according
to simple arguments, half the accretion luminosity of CVs ought to be
emitted in the X-ray band.  Unfortunately, this expectation has not
been fulfilled observationally in most cases --- we must be missing
something, either theoretically or observationally. 

It is an exciting era in X-ray astronomy, with the recent launches
of \chandra\ and \xmm, and the upcoming launch of \astroe.  For example,
not only is the \chandra\ image of the Galactic Center region breath-taking,
a significant fraction of the thousands of point sources may be (magnetic)
CVs \citep{Mea2003}.

In this review, I will define ``X-rays'' as the range 0.1--25 keV,
or roughly 125--0.5\AA.  This is a factor of 250 in wavelength, equivalent
to 1000\AA\ (UV) to 25 $\mu$m (mid-IR).  It is therefore not surprising to find
several distinct spectral components that contribute to the X-ray emission
from CVs.  Single X-ray instruments often cover a factor of 25 in wavelength;
this is a strength of X-ray astronomy.  On the other hand, ``high resolution
spectroscopy'' is a relative term.  In terms of spectral resolution,
\rosat\ PSPC data are comparable to UBVRI photometry.  CCD-based instruments
(\asca\ SIS, \chandra\ ACIS, and \xmm\ EPIC) have typical resolution at 6
keV of 2\%, and therefore may be compared to narrow-band photometry
($\Delta\lambda\sim$100\AA\ at 5000\AA, for example).  Of the existing
instruments, only the \chandra\ and \xmm\ gratings can deliver what
optical astronomers would consider spectroscopy on a routine basis.

This review includes my personal selection of highlights from recent
X-ray observations, with a natural emphasis on the \chandra\ and
\xmm\ results.  I will emphasize magnetic CVs, but will include some
results on non-magnetic CVs.  I will, however, not include SSSs,
novae in outburst, nova shells, or globular cluster sources.  Readers
are referred to other papers in this volume for these subjects.

As a starting point for discussion, I will adopt the following as
the baseline models for X-ray emission in non-magnetic and magnetic
CVs.  In the former, the Keplerian accretion disk is connected to
the white dwarf surface via a boundary layer.  It is optically thin
and emits hard X-rays in low accretion rate systems \citep{PR1985a}.
In high accretion rate systems, the core of the boundary layer is
optically thick and emits soft X-rays, while some optically thin hard
X-rays remain \citep{PR1985b}. In magnetic systems, accretion is in
the form of columns that are nearly vertical.  Hard X-rays from the
post-shock plasma is usually dominant in the IPs \citep{P1994},
while soft X-rays from the photosphere is the most prominent
feature of polars \citep{C1990}.

\section{Polars: Low State and Energy Balance}

The \xmm\ survey of polars have revealed the detailed pictures of
X-ray emissions in a substantial number of polars.  Moreover, the
survey has produced two notable results on the polars as a class.
One is that a large fraction (16 out of 37 surveyed) is seen to be
in a low state during the \xmm\ observations \citep{Rea2004}.  This
has an obvious impact on the efficiency of X-ray surveys to detect
polars, and hence their space density.

Another result is on the soft X-ray/hard X-ray luminosity ratio
of polars.  There is a long history of controversy
over both what ratios to expect theoretically, and what ratios observations
actually indicate.  In this latest contribution on this subject,
\citet{RC2004} conclude that the majority of systems have a soft to
hard luminosity ratios that are consistent with a pure irradiation origin
of the soft component.  However, a significant minority shows a ``soft
excess,'' indicating direct, mechanical heating of the white dwarf atmosphere.
Compared to previous efforts, this study has the advantage of far superior
data on the hard component: \xmm\ is the first imaging instrument capable
of producing high quality spectrum of the hard component on more than
a handful of polars, over the range 0.5--10 keV.  \rosat\ PSPC, in
contrast, could only detect the soft end of this component.  Moreover,
\citet{RC2004} utilize a sophisticated spectral model to fit the
hard component spectrum, rather than relying on a simple Bremsstrahlung
model often with a fixed temperature.

Nevertheless, there is room for further investigation into this issue.
One particular question is the true spectral shape of the soft component.
As \cite{M1999} shows using \euve\ data, the choice of blackbody
vs. stellar atmosphere models can greatly change the inferred bolometric
luminosity.  Moreover, the degree of departure of the spectrum from
a pure blackbody is an indicator of the origin of the soft component,
since strong irradiation causes temperature inversion and therefore
weakens atmospheric features.  Do the polars with no soft excess
possess more blackbody-like soft component that those with strong
excess?  We can hope to learn a great deal on this subject over the
next several years through observations and analysis of \chandra\ LETG
data of polars.

\section{IPs: X-ray Confirmations and Broad Band Spectra}

Coherent X-ray pulses at periods significantly shorter than the
orbital period (P$_{\rm orb}$) is a defining characteristic of IPs.
De Martino and colleagues have been conducting a systematic campaign
to observe IPs and IP candidates in the \sax, \xte, and now \xmm, with
the X-ray confirmation of their IP nature as an important initial
objective.  A latest example can be found in \citet{dMea2005},
in which \xmm\ observation of HT~Cam (P$_{\rm orb}$=86 min)
has revealed a coherent X-ray modulation at the optically-identified
spin period of 515 s.  We now have three X-ray confirmed IPs below the
period gap (EX~Hya, V1025~Cen, and HT~Cam), while a fourth system,
DW~Cnc, almost certainly belongs in this group but still awaits
X-ray confirmation.  These systems raise an interesting evolutionary
question of if and when they might synchronize to become polars.

In addition, three of the four systems (though not HT~Cam) have
a long ($\gg$0.1P$_{\rm orb}$) spin period and should not be able to
form a true Keplerian accretion disk, yet there are observational
signatures of axisymmetric, disk-like structures.  This conundrum
may have been solved by the recent numerical simulations, in which
a new regime of diamagnetic blob accretion has been found \citep{Nea2004}.
In this regime, accretion is fed from a ring-like structure near the
outer edge of the white dwarf Roche-lobe, and the predicted equilibrium
spin period is indeed long ($\gg$0.1P$_{\rm orb}$).

Returning to HT~Cam, there is little sign of significant photoelectric
absorption in its X-ray spectrum, both in the average and in the phase
resolved spectra \citep{dMea2005}.  In this, HT~Cam is similar to EX~Hya,
V1025~Cen, and YY~Dra.  It is possible that this is the hallmark of a
low accretion rate IPs.

The case of WX~Pyx illustrates the potential of pointed X-ray observations
for serendipitous discoveries.  This hydrogen rich CV was discovered
serendipitously in the \einstein\ observation of the nearby galaxy,
NGC~2613.  The discovery of a stable 26 min period \citep{Oea1996},
combined with the presence of hydrogen (therefore P$_{\rm orb} >$ 70 min),
indicated this system is an IP.  The X-ray confirmation of this
classification has been obtained through \xmm\ observations of NGC~2613
(Mukai, de Martino et al., in preparation).  Among the operational X-ray
satellites of today, \xmm\ has a relatively large field of view, and so
has a great potential for serendipitous discoveries (many CVs among them,
one hopes), particularly because the project has a heavy investment in the
\xmm\ serendipitous X-ray sky survey.

\section{Eclipses in Non-magnetic CVs}

\xmm\ is making important contributions to the studies of X-ray
eclipses in non-magnetic CVs, because it has the largest effective
area among imaging X-ray telescopes.  In quiescent dwarf novae,
the existence of X-ray eclipses has been established through
\rosat\ and \asca\ observations (see, e.g., \citealt{Mea1997}),
and the X-ray emission region has been known to be compact.
However, how compact, and is it really consistent with a boundary
layer?  The recent re-analysis of \xmm\ observation of OY~Car
\citep{WW2003} show that, surprisingly, the X-ray eclipse is
narrower than in the optical.  Since the second and the third
contacts align between the optical and X-ray data, while the first
and the fourth contacts do not, a vertical displacement between
the emitting regions is required.  Can this be accommodated within the
boundary layer picture?  Or does this require a significant magnetic
field in OY~Car?  If the latter is the case, the magnetic poles
cannot have a significant optical emission, even though the upper
pole must dominate the observed X-rays.  This subject is likely
to see further developments in the near future based on
\xmm\ observations of Z~Cha and HT~Cas.

\begin{figure}[!ht]
\plotfiddle{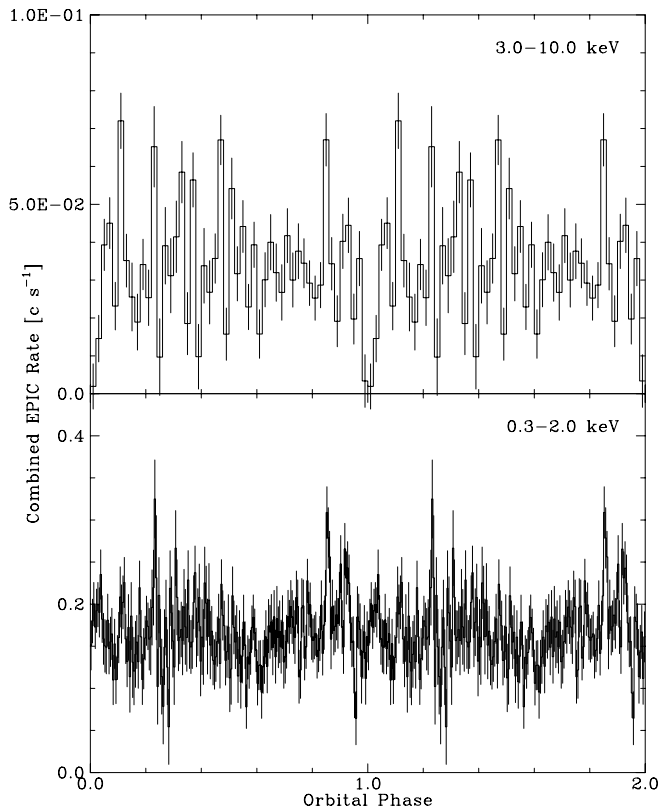}{11 cm}{0.0}{50.0}{50.0}{-168}{-36}
\caption{Folded X-ray light curves of UX~UMa in two energy band.
There is a hard X-ray eclipse, but not in the soft X-rays.
Taken from \citet{Pea2004}.}
\label{uxuma}
\end{figure}

In contrast, in high accretion rate systems (dwarf novae in outburst
and in nova-like systems), soft X-ray eclipses have never been observed.
This includes the bright nova-like system, UX~UMa, which is deeply
eclipsing in the optical and the UV.  It has now been shown that
there is an X-ray eclipse, but only above $\sim$ 2 keV (Figure\,\ref{uxuma};
\citealt{Pea2004}).  Spectrally, the hard component is strongly absorbed,
and contributes few counts below 2 keV.  This presumably originates
in the boundary layer.  The origin of the soft component is more problematic.
\citet{Pea2004} have speculated that this component may be due to scattering
in the accretion disk wind.  This certainly has the right size scale,
but are there enough E$\sim$1 keV photons that can be scattered?
\citet{W2005} explores a possible alternative origin in connection with
the outburst observations of WZ~Sge.

\section{WZ Sge and Friends: More Magnetic CVs?}

As is well known, WZ~Sge is a dwarf nova with an extremely long
recurrence period.  The standard version of the disk instability
model has difficulties explaining this system \citep{Lea1999}, and
additions and/or modifications have been sought to remedy this.
Another well known puzzle regarding WZ~Sge is the rapid oscillations
seen in the optical and the UV.  If it were not for the fact that
there are two stable periods, one at 27.87 s and the other at 28.96 s,
the magnetic CV interpretation would have been accepted universally.
Even with the multiple periods, the magnetic CV interpretation remains
an attractive explanation for the unusual outburst properties of WZ~Sge.

\begin{figure}[!ht]
\plotfiddle{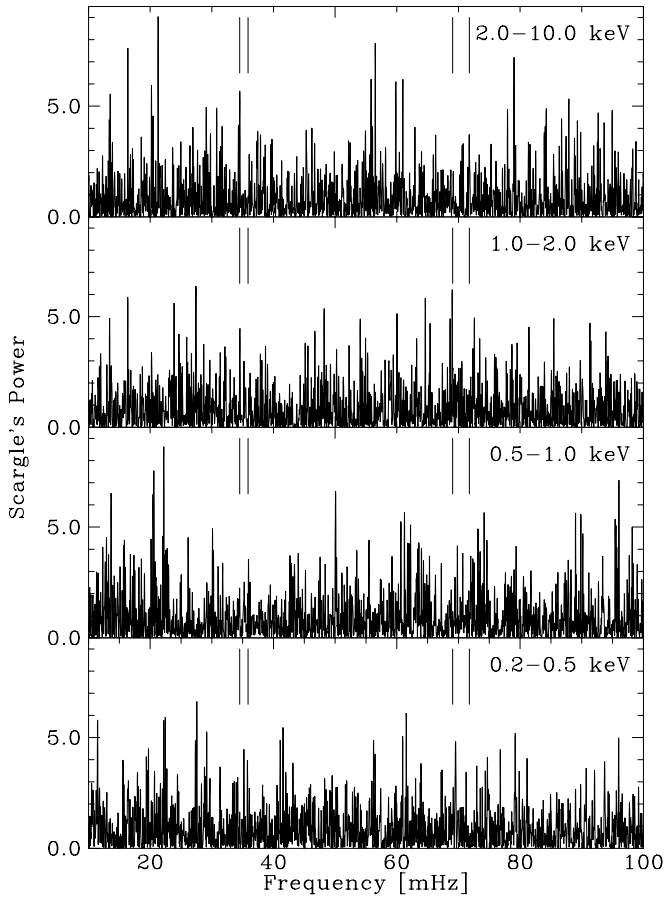}{12.5 cm}{0.0}{50.0}{50.0}{-168}{-12}
\caption{Power spectra of WZ~Sge in quiescence in 4 energy bands,
from the 2003 May \xmm\ observation.  Marked frequencies corresponds
to 28.96 s, 27.87 s, and their first harmonics.}
\label{wzsge}
\end{figure}

\citet{MP2004} present the results from 2003 May \xmm\ observation
of WZ~Sge.  The spectrum can be fit with a multi-temperature plasma
emission model with low absorption; it is similar to those of
non-magnetic CVs and low accretion rate IPs.  The X-ray luminosity
is $\sim 2.5 \times 10^{30}$ ergs\,s$^{-1}$, requiring an accretion
rate of $\sim 5 \times 10^{13}$ g\,s$^{-1}$.  Even though these values
are modest, they still exceed the expectations from the standard
disk instability model.  Most importantly, the 27.87 s signal is not
present in the \xmm\ data (see Figure\,\ref{wzsge}; the detection of
this periods was claimed in an earlier \asca\ data).  The 28.96 s signal
is definitely present in the contemporaneous optical data, and may also
be present at a low level in the \xmm\ data (Figure\,\ref{wzsge}).  Thus,
the quiescent \xmm\ data fail to resolve the issue of the origin of
these periods.

It is important to consider what would qualify a CV as a magnetic
system.  Conceptually, any CV in which the magnetic field of the
white dwarf is strong enough to control the accretion flow is a
magnetic CV\footnote{An interesting complication is that accretion may
lead to the evolution of the magnetic field of the white dwarf \citep{C2002}.}.
It is quite possible that systems traditionally considered ``non-magnetic''
are in fact magnetic.  Observers must then ask what constitutes strong
evidence for this.

As an example, \citet{Bea2004} have analyzed the complete sample of
\asca\ observations of ``non-magnetic'' CVs (29 in all), looking for
periodic modulations and spectra that are indicative of magnetic CVs.
They find 3 candidate magnetic CVs: LS~Peg, EI~UMa, and V426~Oph.
\citet{Hea2004} also revived the magnetic interpretation for V426~Oph
based on the \chandra\ HETG spectroscopy as well as their re-analysis
of the archival X-ray data.  V426~Oph typifies the difficulties
sometimes encountered in searching for new magnetic CVs using X-ray
data.  As \citet{Hea2004} note, previously reported X-ray periods
appear to be transient and quasi-periodic.  The periods identified
by \citet{Bea2004} (29.2 min) and \citet{Hea2004} (4.2 hr) do not
agree with each other.  In summary, while there are good reasons
to propose V426~Oph as an IP candidate, there does not yet appear
to be a compelling evidence for its magnetic nature, either.

\citet{Bea2004} do not find evidence for OY~Car or WZ~Sge
being magnetic in the \asca\ data.  In the \xmm\ data on OY~Car,
in addition to the eclipse timing which may favor a magnetic
interpretation, the presence of a 2240 s period has been found
\citep{Rea2001,HR2004}.  If this really is the spin period,
and the system is in spin equilibrium, this almost certainly
excludes the presence of a normal, Keplerian accretion disk.
Yet OY~Car is a bona fide dwarf nova.  If more dwarf novae than
just WZ~Sge are to be interpreted as magnetic CVs, it will be
necessary to apply the disk instability model with a
white dwarf magnetic field to a wide range of system parameters
to conduct a systematic study of the predicted outburst properties.
This must be compared with the known outburst properties of
WZ~Sge, OY~Car, V426~Oph, as well as the bona fide IPs.

Until this is done, it is prudent to be cautious
of magnetic CV interpretation of dwarf novae, unless the spin period
is detected convincingly in multiple datasets.  In addition,
X-ray spectra that are affected heavily by intrinsic absorption
may also be taken as suggestive of magnetic CVs, but this
again is not definitive.

\section{\chandra\ HETG Spectra}

Although high resolution X-ray spectroscopy is also possible
using \xmm\ RGS and \chandra\ LETG, \chandra\ HETG has
been the most productive instrument thus far for CVs.
Here, then, are some highlights from the HETG.

\subsection{Cooling Flow Type Spectrum in CVs}

\begin{figure}[!ht]
\plotfiddle{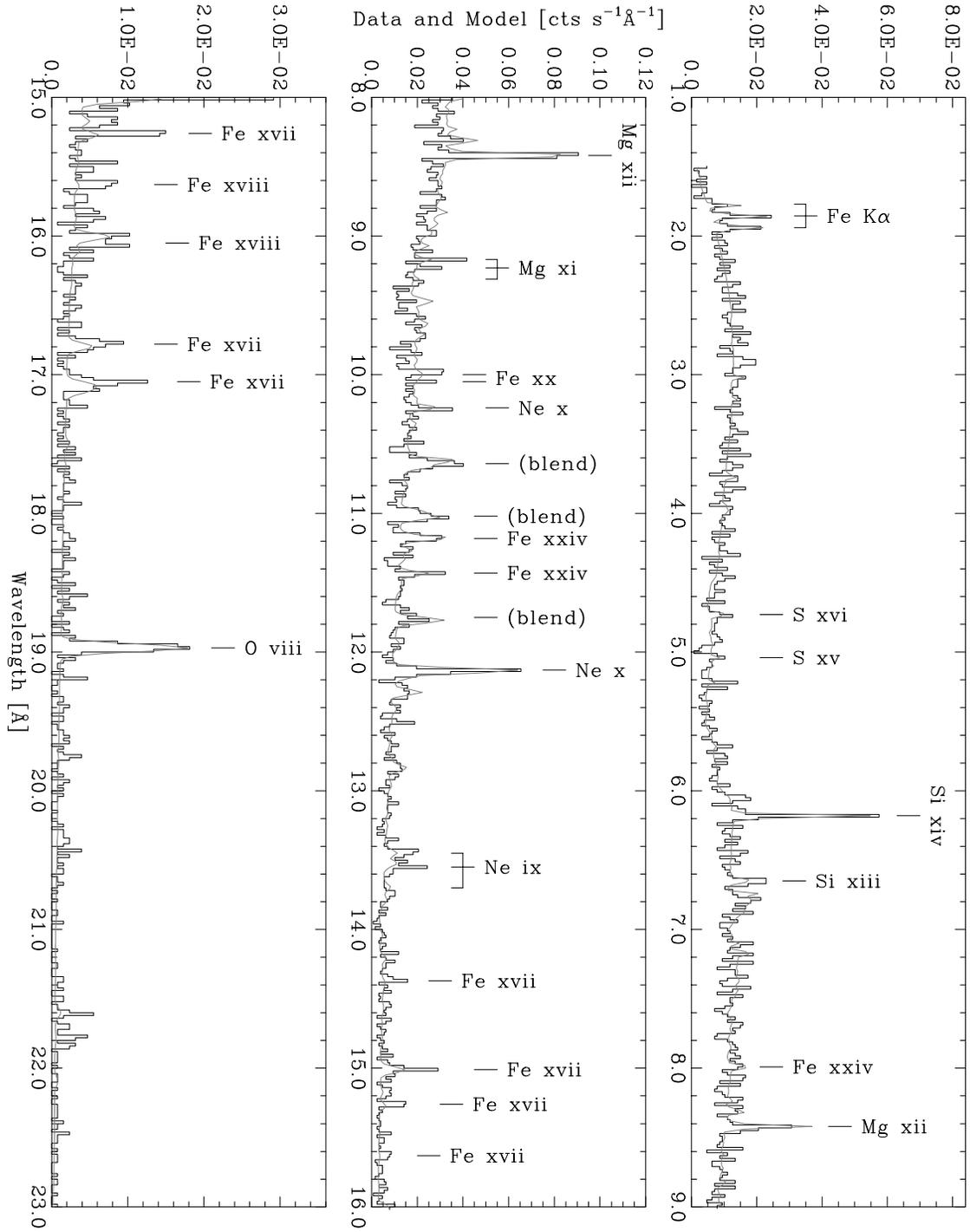}{18.5 cm}{0.0}{90.0}{90.0}{-252}{-54}
\caption{\chandra\ HETG spectrum of V603~Aql, an example of a cooling
flow type spectrum.}
\label{v603aql}
\end{figure}

\citet{twotypes} have compared \chandra\ HETG spectra of seven CVs,
and found two types.  One type (EX~Hya, V603~Aql, U~Gem in quiescence,
and SS~Cyg in quiescence) is found to be consistent with the cooling
flow model, originally developed for clusters of galaxies (see
Figure\,\ref{v603aql} for an example).  This is
a specific version of multi-temperature plasma model with a simple
physical interpretation.  There is little doubt that such multi-temperature,
cooling plasma model is necessary to fit the high resolution, high
quality X-ray spectra of these CVs.  What is less clear is whether
the cooling flow model fit is correct or unique among all the
multi-temperature plasma models.  In fact, physical
considerations suggest there may be subtle differences between real
CVs and the cooling flow model, primarily near both extremes of
temperature distribution due to boundary conditions.  In the case
of non-magnetic CVs, \citet{Pea2003} contains a comparison of
several physical models in the context of \chandra\ HETG data
on WX~Hyi.  The great strength of the cooling flow
model, however, is that it is readily available in spectral fitting
package, {\tt xspec}, and one can obtain estimates of the global
accretion rate and other parameters of interest.  More recently,
\citet{MO2005} have attempted to derive the emission measure
distribution as a function of temperature from \chandra\ HETG
data on V603~Aql.  Although only partially successful, this
approach, which combines global model fits with individual line
fits, may prove fruitful in the future.

It is important to stress that the boundary layer emissions are
expected to be multi-temperature in nature.  Some earlier works treat
the boundary layer emission as that of a single temperature
plasma.  While a theoretical plot of the plasma temperature versus
the height above the white dwarf surface of a boundary layer
shows a large volume of near constant temperature,
this is because this region is low density, cools only
very inefficiently, and is advection-dominated \citep{NP1993}.
The actual X-ray emission mostly comes from a small volume
near the white dwarf surface where the temperature plot
is almost vertical.  In this region, the temperature drops
because energy is radiated away, which in turn is because
of the high density.  Single temperature treatments worked
well with lower quality X-ray data, but are completely
inadequate for \chandra\ HETG data.

Note that \citet{twotypes} did not discuss the spectra
of U~Gem and SS~Cyg in outburst.  There are drastic changes
from quiescence to outburst, and also differences between
U~Gem and SS~Cyg \citep{Mea2005}.  One clear difference
is the greatly increased line width in outburst.  It will be
interesting to investigate whether a cooling flow model, with
large line widths, can be applied to \chandra\ HETG data
of dwarf novae in outburst or not.

\subsection{Photoionized Spectrum in CVs}

\citet{twotypes} have fitted the \chandra\ HETG spectra of
V1223~Sgr, AO~Psc, and GK~Per using a photoionized plasma
model originally developed for active galactic nuclei.
It is important to note that this is a phenomenological
model, in that a power law continuum is assumed without physical
justification.  The main strength of their result is the use of
a physically sophisticated model of photoionized line emission.
The reasons for adopting this model over the cooling flow model
for the low energy emission lines have been articulated by \citet{twotypes}.

So why the power law continuum?  The reason \citet{twotypes} have
adopted the power law prescription is that the observed continua are
too hard to be from a cooling flow. The cooling flow continuum is
a superposition of thermal Bremsstrahlung of many different temperatures,
and resembles a power law of photon index $\sim$1.4.  In contrast, these
three magnetic CVs have a much harder (photon index $\sim$0.5 or
harder) continuum in the \chandra\ HETG range.

However, from higher energy observation (e.g., with \ginga\ and \xte)
we know that these magnetic CVs have thermal Bremsstrahlung like continua
above 10 keV \citep{I1991}.  Complex, phase-dependent absorption
is an essential ingredient in our understanding of the X-ray
spectrum of magnetic CVs \citep{NW1989}.  So, one expects to see
absorbed Bremsstrahlung continua in these systems below 10 keV.

A simple absorber, however, produces an exponential cut-off at low energies.
A superposition of a few discrete partial covering absorbers only
produces a series of exponential cut-offs at different energies.
These will not produce hard power laws as observed in V1223~Sgr,
AO~Psc, and GK~Per with \chandra\ HETG.  However, the real absorber
in magnetic CVs can be more complex than a few discrete partial
covering absorbers.  \citet{DM1998} have explored the absorption
expected from the pre-shock flow.  Our lines of sight to different
parts of the post-shock region pass through different amounts of
pre-shock flow.  Although a full numerical model is the most
accurate, they have also developed an approximation in which the
covering fraction is a power law function of N$_{\rm H}$.  This has
been implemented as the {\tt pwab} model in {\tt xspec}.  Interestingly,
this results in a power-law low energy cut-off, just as required if we
are to explain the \chandra\ HETG continuum as a cooling flow continuum with
complex absorption.

So, the full physical model for V1223~Sgr, AO~Psc, and GK~Per
is likely to be a hybrid.  There is an underlying cooling flow
emission from the post-shock region.  We observe this through
a complex absorber that is the immediate pre-shock flow.  Given
the physical parameters of these CVs, it would be difficult to
produce Fe K$\alpha$ lines through photoionization; they are
far more likely to be intrinsic to the cooling flow.  However,
low energy emission lines are likely to be produced as a
result of photoionization of the pre-shock flow \citep{twotypes}.

Note that, according to \citet{DM1998} and to Rainger (unpublished
work), the varying geometry of the pre-shock complex absorber
is the likely cause of X-ray orbital (=spin) modulation in polars.
Presumably the same model can be applied to the spin modulation in
IPs.  This is distinct from the ``accretion curtain'' absorber
which is much further away, which affects the entire post-shock
emission region simultaneously and hence is likely to produce
a strongly energy dependent dip.

\subsection{Plasma Diagnostics}

A high density is expected in the X-ray emission regions in
magnetic CVs.  For a fiducial local accretion rate of
1 g\,cm$^{-2}$s$^{-1}$, decelerating by a factor of 4 from
a free-fall velocity of 3,000 km\,s$^{-1}$, the expected density
is of order $10^{16}$ cm$^{-3}$ just below the shock, and higher
lower down.  There is far greater uncertainties for non-magnetic CVs,
but somewhat lower densities might be expected.
X-ray spectroscopy has the potential to provide strong constraints
on the density, via determinations of line ratios.  As the He-like
triplet ratios are also affected by UV photo-excitation, Mauche
and colleagues have turned their attention to potential Fe L line
diagnostics, as applied to the high quality \chandra\ HETG spectrum
of EX~Hya.

\citet{fel17} have used the ratio of Fe~XVII lines at 17.10 \AA\ and
at 17.05 \AA.  However, this ratio can also be affected by UV photoexcitation.
\citet{fel22} have used the ratio of Fe~XXII lines at 11.92 \AA\ and
at 11.77 \AA.  They find this ratio to be insensitive to photoexcitation,
and to have a critical density of $\sim 5 \times 10^{13}$ cm$^{-3}$.
The ratio observed in the \chandra\ HETG spectrum of EX~Hya then implies
a density near $1 \times 10^{14}$ cm$^{-3}$ for a 12 million degree
plasma.   These Fe L diagnostics require a high signal-to-noise
data, and also requires the source to be relatively unabsorbed.
Thus application has so far been limited just to EX~Hya.  However,
these and other diagnostics have the potential to allow direct
measurements of the plasma density and confront the physical
models of the X-ray emitting regions in CVs.

The Fe K$\alpha$ lines merit particular attention, not least because
these have been the longest studied X-ray emission lines, due to a
combination of astrophysical and technological reasons.  A couple of
developments merit mention here.  First, \citet{HM2004} have analyzed
the \chandra\ HETG data on five magnetic CVs.  They do not confirm
the earlier result based on \asca\ data that the H-like and He-like
lines of AO~Psc is Compton broadened.  There are, however, hints of
structures within these components.  Second, \citet{Tea2004} have
studied the spin modulations of K$\alpha$ line intensities in
polars and in IPs.  They found significant modulations in polars,
but not in IPs, and explain their findings in terms of resonant
scattering of line photons.

\subsection{X-ray Radial Velocity Studies}

Perhaps the most exciting new result is the detection of X-ray
radial velocity motion in EX~Hya \citep{HBM2004}.  One could
say that X-ray spectroscopy of CVs has finally come of age.
Astrophysically, this is exciting, since optical spectroscopy
generally does not allow an unambiguous measurement of the radial
velocity motion of the white dwarf.  The measured amplitude of
$\sim$58 km\,s$^{-1}$ in EX~Hya implies a white dwarf mass of
$\sim$0.5 M$_\odot$.  Another important result is the lack of
significant spin modulation of radial velocities.

\section{Conclusions and Future Prospects}

Recent X-ray observations of CVs have been very fruitful, as
I hope the preceding sections have shown.  The large effective area of
\xmm\ EPIC allows moderate resolution spectroscopy of relatively
faint CVs, and provides the best X-ray eclipse light curves
of magnetic and non-magnetic CVs.  X-ray confirmations of IP
nature and serendipitous discoveries are other strong areas
for \xmm.  \chandra\ HETG has allowed us to study the
temperature distribution and the densities in some of the
X-ray brightest CVs, and even an X-ray radial velocity study!

What are some of the unanswered questions?

\begin{enumerate}
\item What is the true spectral shape and the true luminosity of
      the soft component in polars?
\item What is the origin of the uneclipsed $\sim$1 keV X-rays in
      UX~UMa?  It is likely this component exists in other
      high accretion rate non-magnetic CVs as well.
\item What is the geometry of hard X-ray emitting regions in
      quiescent dwarf novae?  Is OY~Car typical, if so is there
      such a thing as the boundary layer?
\item Is WZ~Sge magnetic or not?  Are there many magnetic CVs
      hiding among dwarf novae?
\item Why are X-ray lines often narrow in CVs, and only appear
      broad in dwarf novae in outburst?
\end{enumerate}

Both \chandra\ and \xmm\ will continue to provide high quality data
on CVs.  \astroe\ is expected to join these observatories
in 2005.  It is particularly suited to the study of the Fe K$\alpha$
lines, with a resolution of $\sim$6 eV and an effective area of
$>$100 cm$^2$.  Since the Fe K$\alpha$ lines arise from the hottest, least dense,
and least decelerated part of the post-shock region in both
magnetic and non-magnetic CVs, there is rich potential for
kinematic studies of X-ray emitting regions in CVs using
\astroe\ data in the Fe K$\alpha$ region.

\end{document}